\documentclass[review,3p,times,11pt,numbers]{elsarticle}

\usepackage[T1]{fontenc}
\usepackage[utf8]{inputenc}
\usepackage{microtype}
\usepackage{graphicx}
\usepackage{amsmath,amssymb}
\usepackage{booktabs}
\usepackage{url}
\usepackage[hidelinks]{hyperref}
\usepackage[font=small,labelfont=bf]{caption}
\usepackage{placeins}

\graphicspath{{figures/}}
\biboptions{numbers,sort&compress}
\setcitestyle{numbers,round}
\journal{Preprint}

\begin{document}

\begin{frontmatter}

\title{Spatially Organized Hillslope Weakening Localizes Retrogressive Thaw Slumps across the Tibetan Plateau}

\author[label1]{Yuting Yang}
\author[label1]{Gang Mei\corref{cor1}}
\ead{gang.mei@cugb.edu.cn}
\cortext[cor1]{Corresponding author}
\author[label1]{Nengxiong Xu}
\author[label1]{Yongshuang Zhang}
\author[label1,label2]{Jianbing Peng}

\address[label1]{School of Engineering and Technology, China University of Geosciences (Beijing), 100083, Beijing, China}
\address[label2]{School of Geological Engineering and Geomatics, Chang'an University, 710064, Xi'an, China}

\begin{abstract}
Long-term permafrost warming on the Tibetan Plateau is continuously and broadly reshaping cold-region landscapes, and has produced numerous localized and clustered retrogressive thaw slumps (RTSs). However, the question why RTSs are highly localized and clustered under the widespread permafrost warming remains incompletely understood. To understand and explain the RTS localization and clustering, we propose the Hillslope Weakening Field (HWF), which represents the spatial distribution of relative hillslope weakening states jointly formed by multiple environmental conditions, including thermal conditions, moisture and soil properties, topographic setting, permafrost and ground-ice conditions, and land-surface cover. We examined the correspondence between the HWF and RTS spatial selectivity across seven study regions on the Tibetan Plateau. Positive HWF--RTS correspondence was observed in six of the seven regions, with consistently strong relationships in five. In these five regions, RTSs preferentially occurred on more strongly weakened hillslopes, with occurrence rates 3.35--5.09 times the regional background rate within the 10\% most strongly weakened slope units. Moreover, high-weakening hillslopes were not spatially isolated but correlated and connected across the landscape. RTS-bearing slope units and their surrounding hillslopes showed higher weakening levels than RTS-unrecorded slope units and their surroundings, and the weakening contrast declined with distance. These results suggest that RTSs preferentially occur on more strongly weakened hillslopes and cluster where high-weakening states are spatially continuous and connected. Our findings show how spatial differences in hillslope weakening states can help explain why geomorphic responses to widespread permafrost warming are highly heterogeneous across permafrost landscapes.
\end{abstract}


\end{frontmatter}

\section{Introduction}\label{sec:1}

Permafrost warming and degradation are reshaping geomorphic processes across cold regions \citep{ref01,ref02}. Retrogressive thaw slumps (RTSs) are rapid hillslope mass-wasting features initiated by the thaw of ice-rich permafrost and represent one of the most pronounced geomorphic responses to permafrost degradation \citep{ref03}. When thermal disturbance, erosion, or other geomorphic processes expose and thaw ground ice or ice-rich permafrost, the resulting thawed material can collapse, slide, and flow downslope \citep{ref04}. Continued headwall retreat then allows the slump to expand progressively. RTSs can rapidly alter hillslope morphology, mobilize large volumes of permafrost material, sediment, and organic carbon, and affect downstream hydrology, water quality, vegetation recovery, and ecosystem carbon cycling \citep{ref03,ref05,ref06,ref07}. In recent years, regional inventories and multitemporal remote-sensing observations have documented increasing numbers of RTSs, expansion of affected areas, and sustained RTS activity across the Tibetan Plateau, making RTSs an important geomorphic expression of ongoing permafrost degradation \citep{ref08,ref09,ref10}.

\begin{figure}[htbp]
\centering
\includegraphics[width=0.88\textwidth,height=0.68\textheight,keepaspectratio]{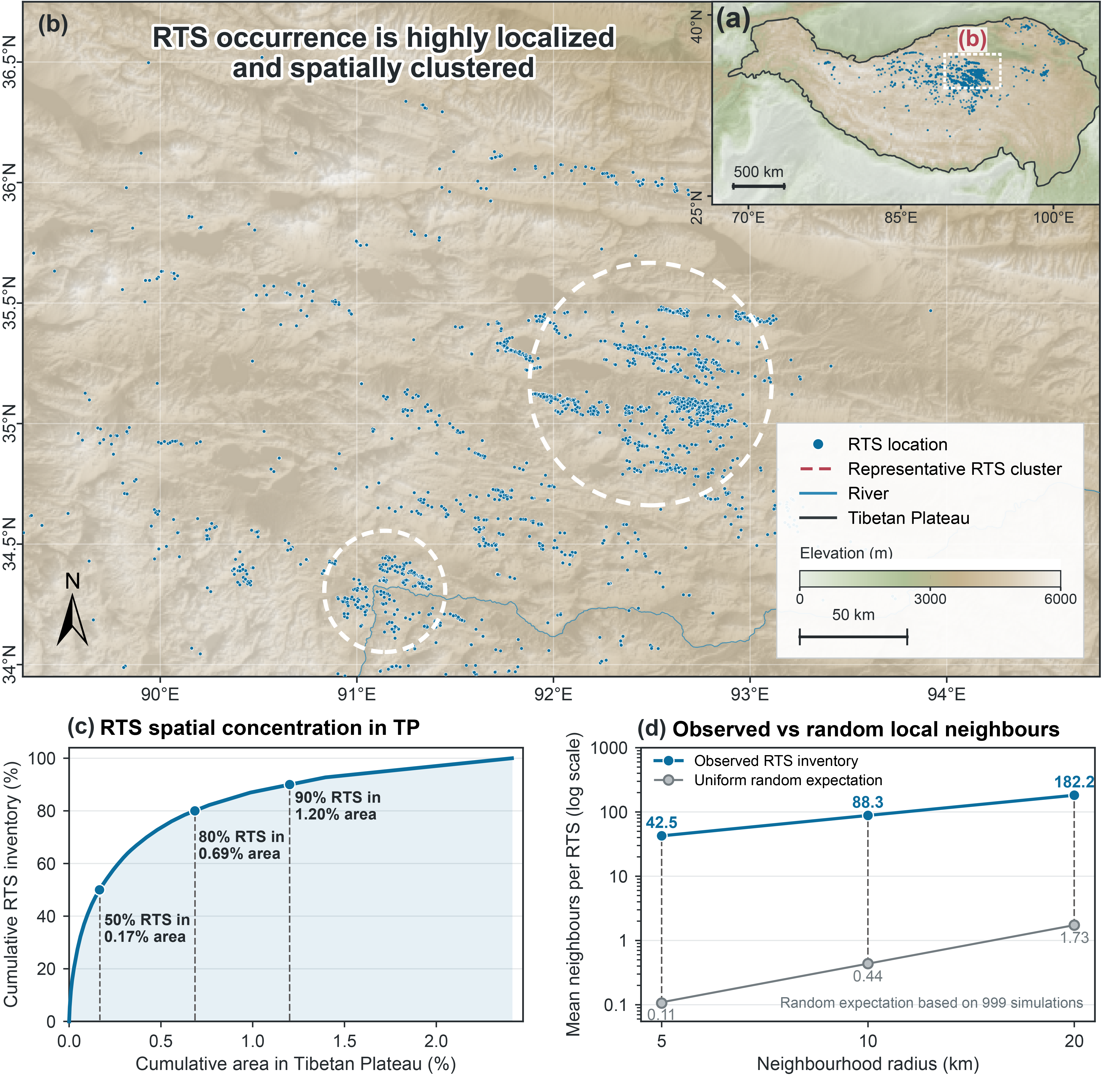}

\caption{Spatial localization and clustering of retrogressive thaw slumps across the Tibetan Plateau. a, Distribution of retrogressive thaw slumps (RTSs) across the Tibetan Plateau. The white dashed rectangle indicates the area enlarged in b. b, Spatial distribution of RTSs within the enlarged area. The white dashed circles indicate two representative high-density clusters. c, Cumulative spatial concentration of RTSs across the Tibetan Plateau. The plateau was divided into 10 $\times$ 10 km equal-area grid cells, which were ranked by the number of RTSs they contained. The highest-ranked grid cells, accounting for 0.17\%, 0.69\%, and 1.20\% of the plateau area, contain 50\%, 80\%, and 90\% of all RTSs in the inventory, respectively. d, Mean number of neighboring RTSs within 5-, 10-, and 20-km neighborhoods around each RTS, compared with the expectation under complete spatial randomness. Blue points show the observed mean numbers of neighboring RTSs, and gray points show the mean values from 999 uniform random simulations across the Tibetan Plateau. The y axis is shown on a logarithmic scale. Together, c and d show that RTSs occupy only a small proportion of the Tibetan Plateau and exhibit strong local spatial clustering.}
\label{fig:1}
\end{figure}

However, a defining spatial feature of RTSs is their strong localization and clustering rather than uniform expansion across warming permafrost landscapes \citep{ref04,ref08,ref11}. Large numbers of RTSs are concentrated within a limited set of small catchments and local areas, while adjacent hillslopes often differ markedly in RTS occurrence. A plateau-wide RTS inventory further shows that most RTSs are concentrated within only a small proportion of the Tibetan Plateau and are much more strongly clustered at local scales than expected from a random spatial distribution (Fig.~\ref{fig:1}). This pattern reveals a clear spatial contrast: permafrost warming and degradation are widespread, whereas RTS development is confined to relatively limited areas and forms spatial clusters. Regional warming, permafrost degradation, and the presence of ground ice provide a broad environmental basis for RTS development \citep{ref03,ref04}, but these conditions do not explain why RTSs develop on some hillslopes but not on adjacent ones, or why they form pronounced spatial clusters. This contrast raises an interesting question: why do retrogressive thaw slumps remain highly localized and clustered under widespread permafrost warming?

Previous studies have identified multiple environmental controls on RTS development, including thermal conditions, moisture conditions, topography, ground ice, and land cover. These factors influence permafrost thermal stability \citep{ref03,ref12}, ground-ice thaw \citep{ref13,ref14}, hillslope water convergence and drainage \citep{ref03,ref05,ref15}, and material mobilization and gravitational forcing \citep{ref15}. Together, these factors define the broad environmental conditions conducive to RTS development. Substantial progress has been made in identifying individual environmental controls on RTS development. However, how these conditions together give rise to the strong spatial selectivity of RTSs remains unresolved. In particular, it is still unclear why RTSs occur on only a small proportion of hillslopes and form pronounced spatial clusters under widespread permafrost warming and degradation.

RTS localization and spatial clustering cannot be readily explained by any single environmental factor because relationships between environmental conditions and hillslope weakening can be nonlinear and region dependent \citep{ref09,ref16}. The same environmental variable may therefore show different response forms and strengths among regions. Moreover, thermal, moisture, topographic, permafrost, and ground-ice conditions operate jointly at the hillslope scale. RTS spatial selectivity may therefore be better understood in relation to hillslope states jointly shaped by multiple environmental conditions than to individual environmental conditions considered separately.

Existing studies have largely treated hillslopes as independent analytical units and quantified relationships between local environmental attributes and RTS occurrence \citep{ref10,ref15,ref17}. However, it remains unclear whether hillslope states jointly shaped by multiple environmental conditions exhibit characteristic spatial scales, continuity, and connectivity across landscapes. It is also unclear whether these spatial properties help explain RTS spatial selectivity, expressed by the concentration of RTSs on only a limited subset of hillslopes and their pronounced spatial clustering. These unresolved questions limit our understanding of how widespread changes in permafrost environments are translated into spatially selective geomorphic responses across hillslopes.

More studies of hillslope instability have suggested that the spatial pattern of geomorphic failure depends not only on local environmental conditions, but also on the magnitude and spatial scale of heterogeneity in hillslope strength. Spatially continuous low-strength zones can influence where failures initiate, how they propagate, and how individual failures merge, thereby shaping the resulting geomorphic pattern \citep{ref18}. This understanding suggests that relationships with individual environmental conditions alone may be insufficient to explain RTS spatial localization and clustering. The hillslope states jointly shaped by multiple environmental conditions, and the spatial properties of those states, also need to be considered.

To understand and explain the RTS localization and clustering, we propose the Hillslope Weakening Field (HWF), which represents the spatial distribution of relative hillslope weakening states jointly formed by multiple environmental conditions, including thermal conditions, moisture and soil properties, topographic setting, permafrost and ground-ice conditions, and land-surface cover. We further hypothesize that these weakening states exhibit continuity, characteristic spatial scales, and connectivity, producing spatially coherent high-weakening domains rather than isolated high-scoring slope units. Under this hypothesis, RTS localization and clustering should correspond not only to higher weakening levels on individual hillslopes but also to the continuity and connectivity of high-weakening states across the landscape.

\begin{figure}[htbp]
\centering
\includegraphics[width=0.98\textwidth,height=0.68\textheight,keepaspectratio]{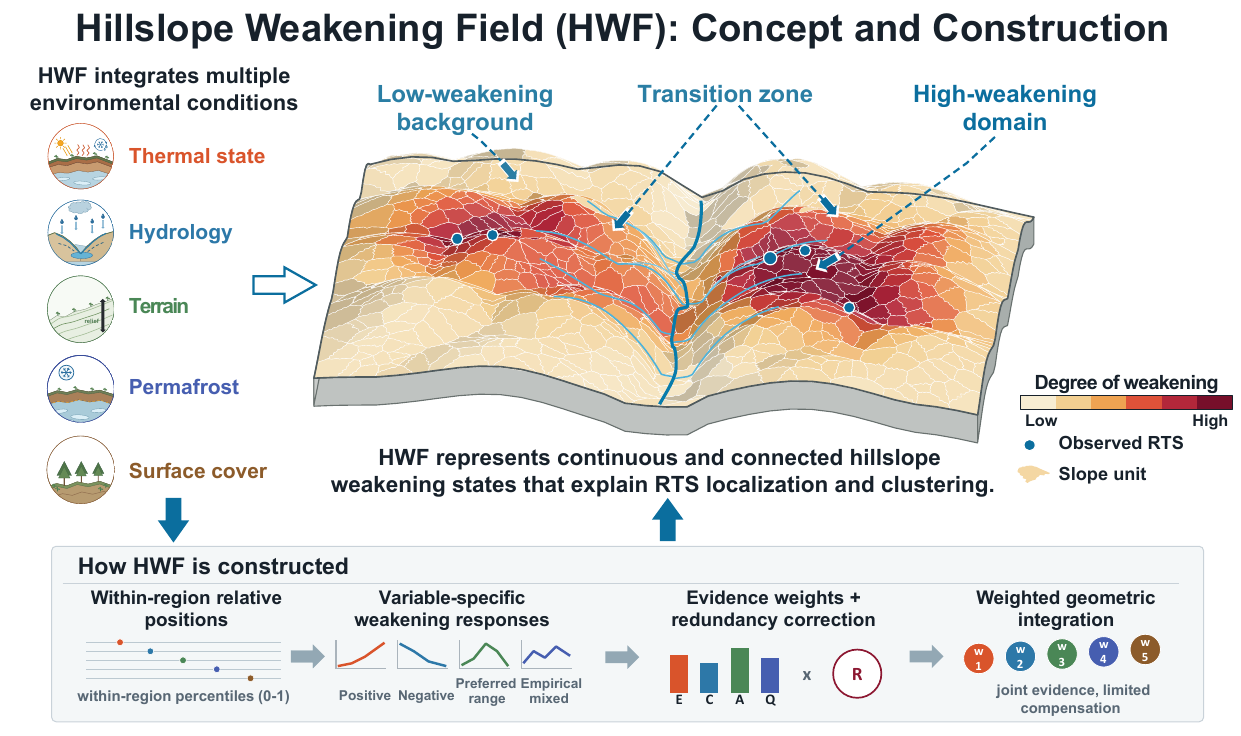}

\caption{Concept and construction of the Hillslope Weakening Field (HWF).}
\label{fig:2}
\end{figure}

\begin{figure}[htbp]
\centering
\includegraphics[width=0.88\textwidth,height=0.72\textheight,keepaspectratio]{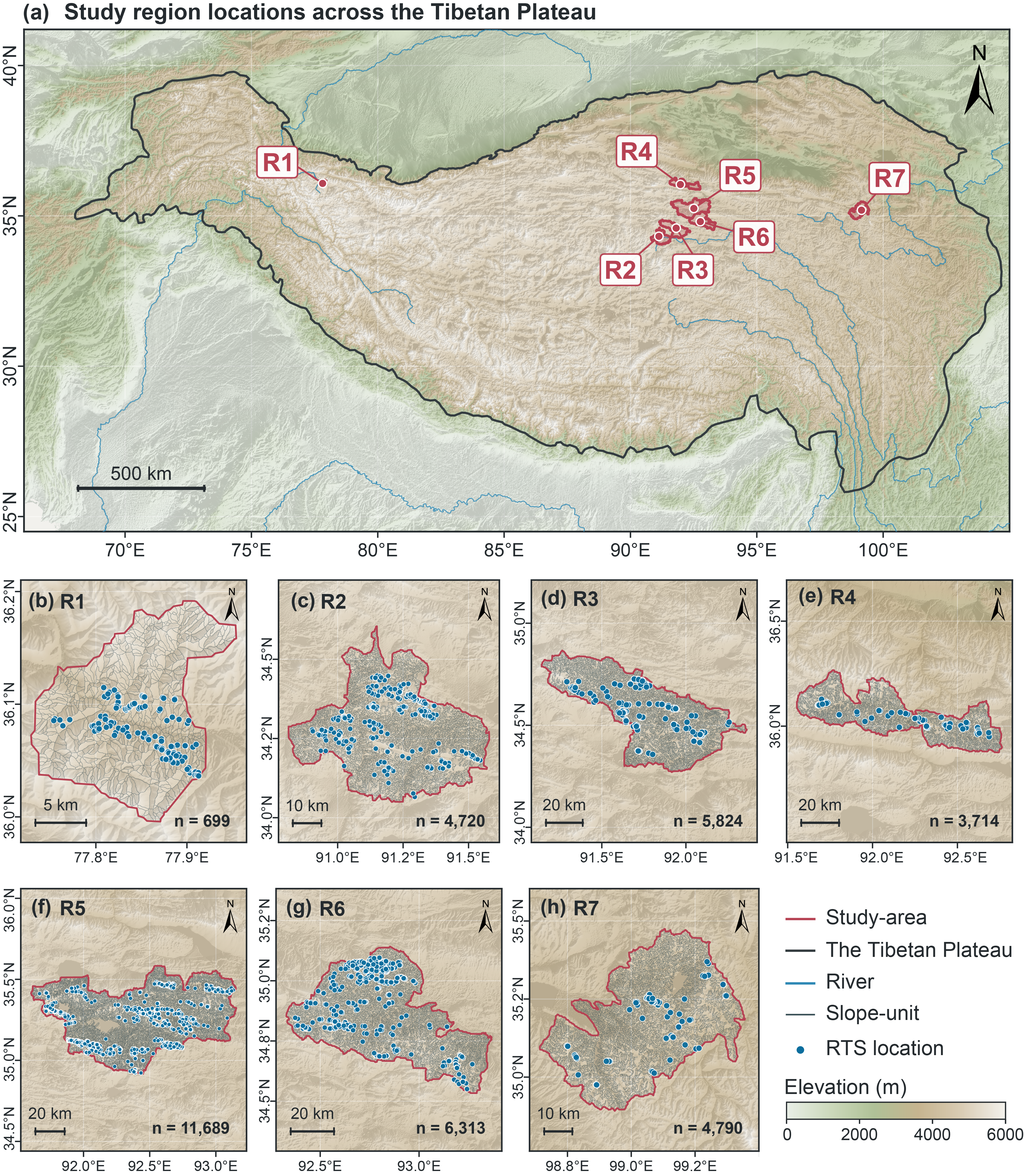}

\caption{Locations of the study regions and distributions of RTSs and slope units.}
\label{fig:3}
\end{figure}

We evaluated this hypothesis across seven study regions on the Tibetan Plateau with contrasting environmental settings (Fig.~\ref{fig:3}). We used slope units that preserve complete hillslope morphology and hillslope--channel connectivity as the basic analytical units and integrated multiple sources of environmental information, including thermal state, hydrological and soil conditions, topography, permafrost and ground ice, and land cover, to construct the HWF. We adopted a complete leave-one-region-out design. For each independent evaluation region, the weakening-response forms and relative evidence weights of environmental factors were determined using the other six regions. RTS records from the evaluation region was therefore not used to construct its HWF and was used only after HWF construction to evaluate HWF--RTS correspondence.

We first examined RTS spatial localization and clustering, thereby defining the spatial pattern to be explained. We then tested, without using RTS records, whether the HWF exhibits spatial autocorrelation and whether high-weakening hillslopes form continuous and connected spatial structures across the landscape. We further examined whether RTSs preferentially occur on hillslopes with higher weakening scores in the HWF, and whether RTS localization and clustering correspond to spatially extensive and locally continuous high-weakening domains and elevated weakening in their surrounding neighborhoods. Finally, we compared the seven regions to assess whether these HWF--RTS relationships recur across different environmental settings and how regional differences relate to environmental conditions. Together, these analyses test whether RTS localization and clustering are associated with both the degree and spatial properties of hillslope weakening states, and whether this correspondence recurs across different permafrost landscapes.

\section{Results}\label{sec:2}

\subsection{RTS distribution is characterized by spatial localization and clustering}\label{sec:2-1}

\begin{figure}[htbp]
\centering
\includegraphics[width=0.98\textwidth,height=0.66\textheight,keepaspectratio]{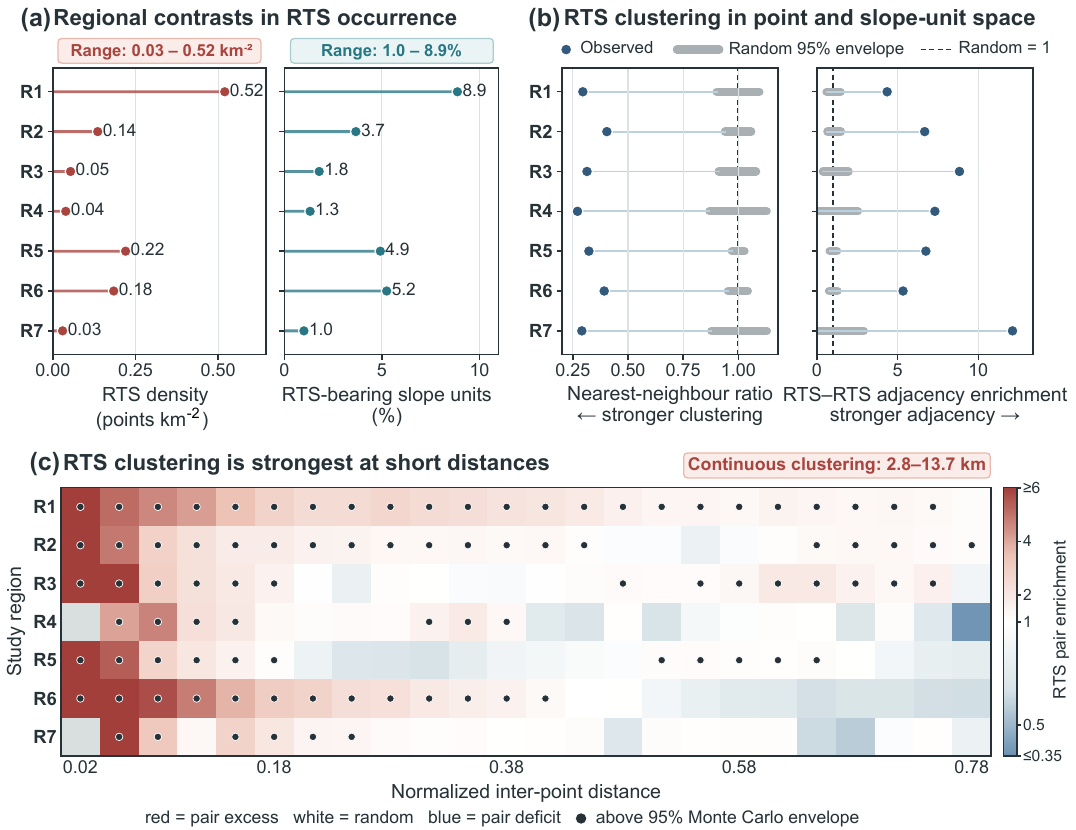}

\caption{RTS distribution is characterized by spatial localization and clustering across all seven study regions. a, RTS density and the proportion of RTS-bearing slope units across the seven study regions. RTS density ranged from 0.03 to 0.52 km$^{-2}$, whereas the proportion of RTS-bearing slope units ranged from 1.0\% to 8.9\%. b, Mean nearest-neighbor distance ratio and RTS--RTS slope-unit adjacency enrichment. Blue points show the observed values, gray lines indicate the corresponding 95\% ranges from randomizations, and dashed lines indicate the random expectation (ratio = 1). Both metrics indicate spatial clustering in all seven regions (one-sided Monte Carlo P = 0.001). c, Enrichment of RTS pairs across normalized distance bins, calculated as the observed number of RTS pairs divided by the mean number of pairs from random patterns within the same distance bin. Red indicates more RTS pairs than randomly expected, white indicates values close to the random expectation, and blue indicates fewer pairs than randomly expected. Black points indicate distance bins in which the observed pair count exceeds the 97.5th percentile of the random results, and gray cells indicate bins with an expected pair count <5. Significant pair enrichment extended continuously from short distances to approximately 2.8--13.7 km across the seven regions.}
\label{fig:4}
\end{figure}

We selected seven study regions on the Tibetan Plateau, delineated using level-12 HydroBASINS subbasin boundaries \citep{ref19}, and used slope units that preserve complete hillslope morphology and hillslope--channel connectivity as the basic analytical units. RTS distribution was strongly localized, while occurrence frequency varied markedly among regions. Among 37,749 slope units, only 1,342 contained at least one RTS, accounting for 3.56\% of all slope units, whereas the remaining 96.44\% contained no recorded RTSs. RTS density varied markedly among regions, ranging from 0.03 km$^{-2}$ in R7 to 0.52 km$^{-2}$ in R1. The proportion of RTS-bearing slope units varied markedly among regions, from 1.0\% in R7 and 1.3\% in R4 to 8.9\% in R1, with intermediate values of 4.9\% and 5.2\% in R5 and R6, respectively (Fig.~\ref{fig:4}a). Thus, RTSs occupied only a small proportion of hillslopes, while their regional occurrence frequency varied markedly.

Nearest-neighbor analysis showed consistent nonrandom clustering of RTS representative points across all seven regions. The observed mean nearest-neighbor distances were only 27.1--40.4\% of the corresponding random expectations and were lower than the 95\% simulation envelopes from 999 region-constrained randomizations in every region (one-sided P = 0.001; Fig.~\ref{fig:4}b, left). In other words, observed nearest-neighbor distances were approximately 60--73\% shorter than expected under spatial randomness. Although RTS occurrence frequency differed markedly among regions, RTSs consistently occurred within limited spatial areas rather than being independently and uniformly distributed across each region.

The same clustering pattern was also evident in the adjacency of RTS-bearing slope units. When slope-unit adjacency was defined by shared boundaries, the observed numbers of adjacent RTS--RTS slope-unit pairs were 4.35 to 12.12 times the area-stratified random expectations and exceeded the corresponding random envelopes in all seven regions (one-sided P = 0.001; Fig.~\ref{fig:4}b, right). Even in R4 and R7, where RTS densities and the proportions of RTS-bearing slope units were relatively low, the observed adjacency among RTS-bearing slope units was 7.31 and 12.12 times the corresponding random expectations, respectively. RTS clustering therefore cannot be explained simply by higher regional occurrence frequency, but is also reflected in enhanced local adjacency among RTS-bearing hillslopes.

Distance-based analysis further revealed the spatial extent of RTS clustering. Across all seven regions, the RTS pair-enrichment analysis showed enrichment above random expectations at short normalized distances, followed by an overall decline toward random levels with increasing distance (Fig.~\ref{fig:4}c). The clustering ranges identified from continuous intervals of significant pair enrichment spanned 2.8 to 13.7 km. The shortest clustering range occurred in R7, whereas clustering in R2 and R6 extended to 12.4 and 13.7 km (Fig.~\ref{fig:4}c and Supplementary Fig. S6). RTS clustering was therefore not confined to individual slope units or directly adjacent hillslopes, but extended to broader neighborhoods comprising multiple hillslopes. The spatial extent and distance-decay pattern of clustering also varied among regions.

Regional occurrence frequency, nearest-neighbor relationships, slope-unit adjacency, and distance-based pair analyses reveal two related spatial characteristics of RTSs. First, RTSs occupy only a limited proportion of hillslopes, indicating strong spatial localization. Second, RTS-bearing hillslopes are not randomly distributed, but show clustering both through enhanced local adjacency and across broader neighborhood distances. Although RTS occurrence frequency differs markedly among regions, this combined pattern of localization and clustering is consistent across all seven study regions. These results define the spatial pattern examined in the following sections: RTSs are localized on a limited subset of hillslopes and further form clusters that extend over characteristic spatial distances. We next examine whether the HWF, jointly shaped by multiple environmental conditions, also exhibits clear spatial autocorrelation and continuous high-weakening domains across the landscape.

\subsection{HWF exhibits spatially correlated and connected high-weakening domains}\label{sec:2-2}

To characterize the spatial properties of the HWF independently of the RTS distribution, we analyzed the continuous distribution of weakening scores, their spatial autocorrelation and characteristic correlation length, and the connectivity among high-weakening hillslopes.

\begin{figure}[htbp]
\centering
\includegraphics[width=0.98\textwidth,height=0.66\textheight,keepaspectratio]{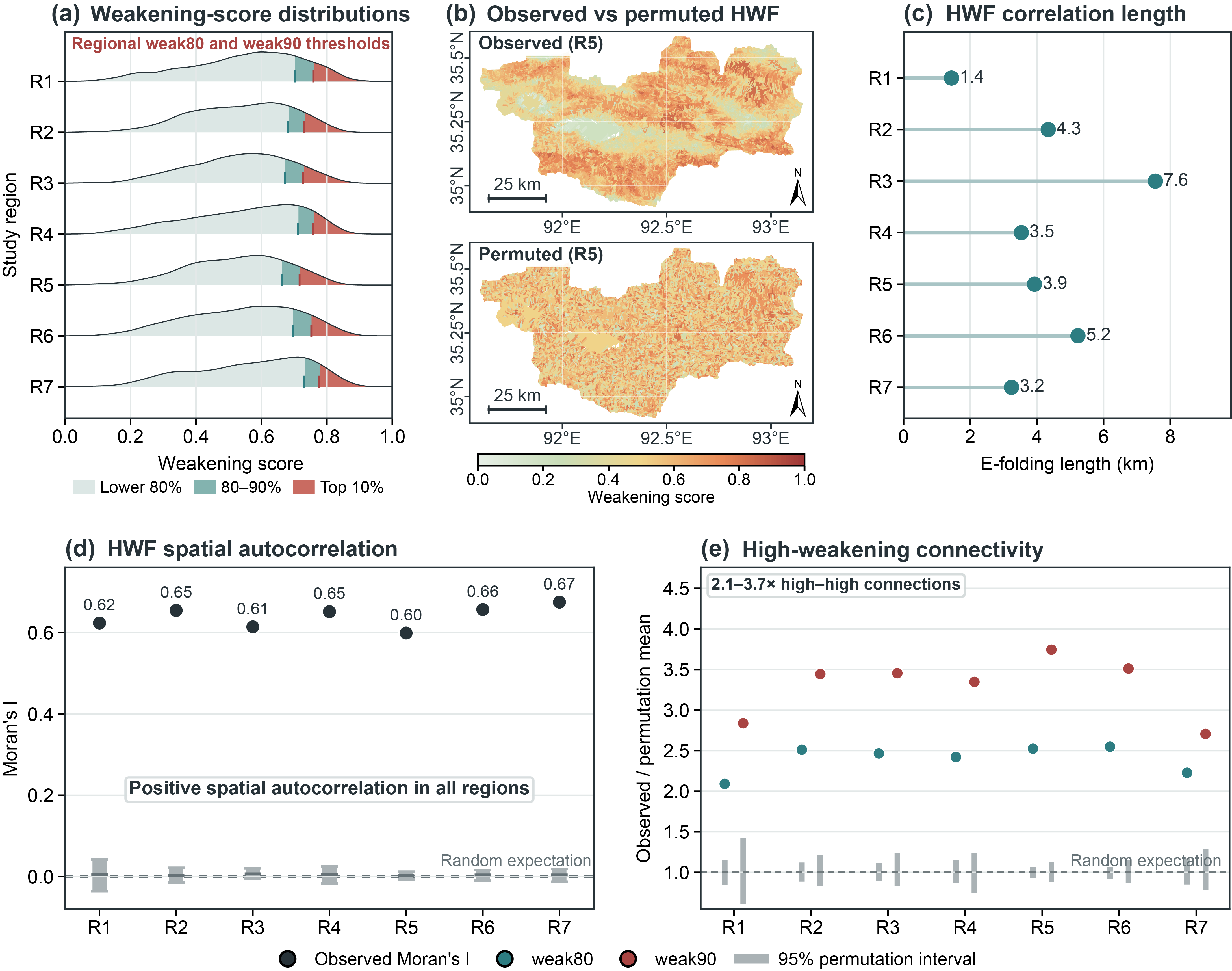}

\caption{HWF exhibits significant spatial correlation and connected high-weakening domains. a, Distribution of weakening scores in the HWF for 37,712 valid slope units obtained through complete leave-one-region-out evaluation. Shaded areas distinguish the lower 80\% of the score distribution and the upper-tail weak80 and weak90 classes in each region, where weak80 and weak90 denote approximately the top 20\% and top 10\% of hillslopes ranked by weakening score, respectively. b, Spatial pattern of weakening scores in the observed HWF in R5 and in one permutation of weakening scores stratified by slope-unit area. The permutation preserves the overall distribution of weakening scores within the region and broadly maintains their relationship with slope-unit area, but rearranges their spatial locations among hillslopes. c, E-folding correlation lengths fitted from the distance-decay relationship of spatial similarity in weakening scores across distance bins, representing the typical spatial extent of HWF continuity. d, Moran's I of the HWF in the seven study regions. Black points indicate observed values, and gray intervals show the 95\% random ranges from 199 weakening-score permutations stratified by slope-unit area. e, Ratios of the numbers of shared-boundary connections among weak80 and weak90 high-weakening hillslopes relative to the mean values from 199 permutations of the same type; gray intervals indicate the 95\% ranges of the permutation results.}
\label{fig:5}
\end{figure}

Weakening scores within the HWF had continuous distributions with clear upper tails in all seven regions (Fig.~\ref{fig:5}a). Within each region, slope units in the top 20\% and top 10\% of weakening scores were defined as weak80 and weak90, respectively, with thresholds of approximately 0.66--0.73 and 0.72--0.78. The weakening scores formed continuous gradients across all regions rather than dividing the landscape into simple weakened and non-weakened classes. High-weakening hillslopes occupied only the upper tail of these distributions.

These high-weakening slope units were not randomly dispersed across the study regions. In R5, for example, the original HWF contained continuous banded and patchy structures of high weakening scores. When the spatial locations of these scores were rearranged while preserving both the overall score distribution and its relationship with slope-unit area, the original high-score structures became markedly fragmented (Fig.~\ref{fig:5}b). This comparison shows that the spatial structure of the HWF depends not only on the weakening state of individual slope units, but also on how different weakening states are arranged among neighboring hillslopes. That is, two spatial fields with the same distribution of weakening scores can differ markedly in their spatial structure.

Spatial autocorrelation analysis further quantified the spatial extent of HWF continuity. Moran's I ranged from 0.60 to 0.67 across the seven regions and was higher than the corresponding area-stratified permutation results in every region (one-sided permutation P = 0.005), indicating significant positive spatial autocorrelation in weakening scores among neighboring and nearby slope units (Fig.~\ref{fig:5}d). Distance-binned standardized spatial autocovariance showed a gradual decline with increasing distance. The estimated e-folding correlation length varied among regions, from 1.4 km in R1 to 7.6 km in R3, while most other regions ranged from approximately 3.2 to 5.2 km (Fig.~\ref{fig:5}c and Supplementary Fig. S7). Thus, HWF continuity extended over characteristic distances that varied among regions.

Consistent spatial connectivity was also evident among high-weakening slope units. In all seven regions, the numbers of shared-boundary connections among weak80 and weak90 slope units exceeded the corresponding area-stratified permutation expectations. Connectivity enrichment ranged from 2.09 to 2.55 for weak80 and from 2.71 to 3.74 for weak90 (Fig.~\ref{fig:5}e). Hillslopes with higher weakening scores therefore tended to occur adjacent to one another rather than being randomly dispersed across the landscape, producing spatially continuous high-weakening areas.

These results show that weakening states within the HWF are spatially correlated rather than randomly dispersed, with characteristic correlation lengths and connected high-weakening structures across the landscape. Having established these spatial properties of the HWF, we next examine whether RTSs preferentially occur on hillslopes with higher weakening scores and whether this spatial preference further extends to high-weakening domains and their surrounding neighborhoods.

\subsection{RTSs preferentially occur on higher-weakening hillslopes within the HWF}\label{sec:2-3}

After characterizing the spatial correlation and connectivity of the HWF, we examined whether RTSs preferentially occurred on hillslopes with higher weakening scores under complete leave-one-region-out evaluation. We conducted seven leave-one-region-out evaluations, with each of R1--R7 serving once as the independent evaluation region and the remaining six regions serving as reference regions. For each evaluation, the weakening-response relationships and relative evidence weights of environmental factors were determined from the six reference regions, and the resulting established HWF construction rules were then applied to the independent evaluation region. RTS records from the independent evaluation region was used only after HWF construction to evaluate the correspondence between RTSs and hillslope weakening states.

\begin{figure}[htbp]
\centering
\includegraphics[width=0.98\textwidth,height=0.63\textheight,keepaspectratio]{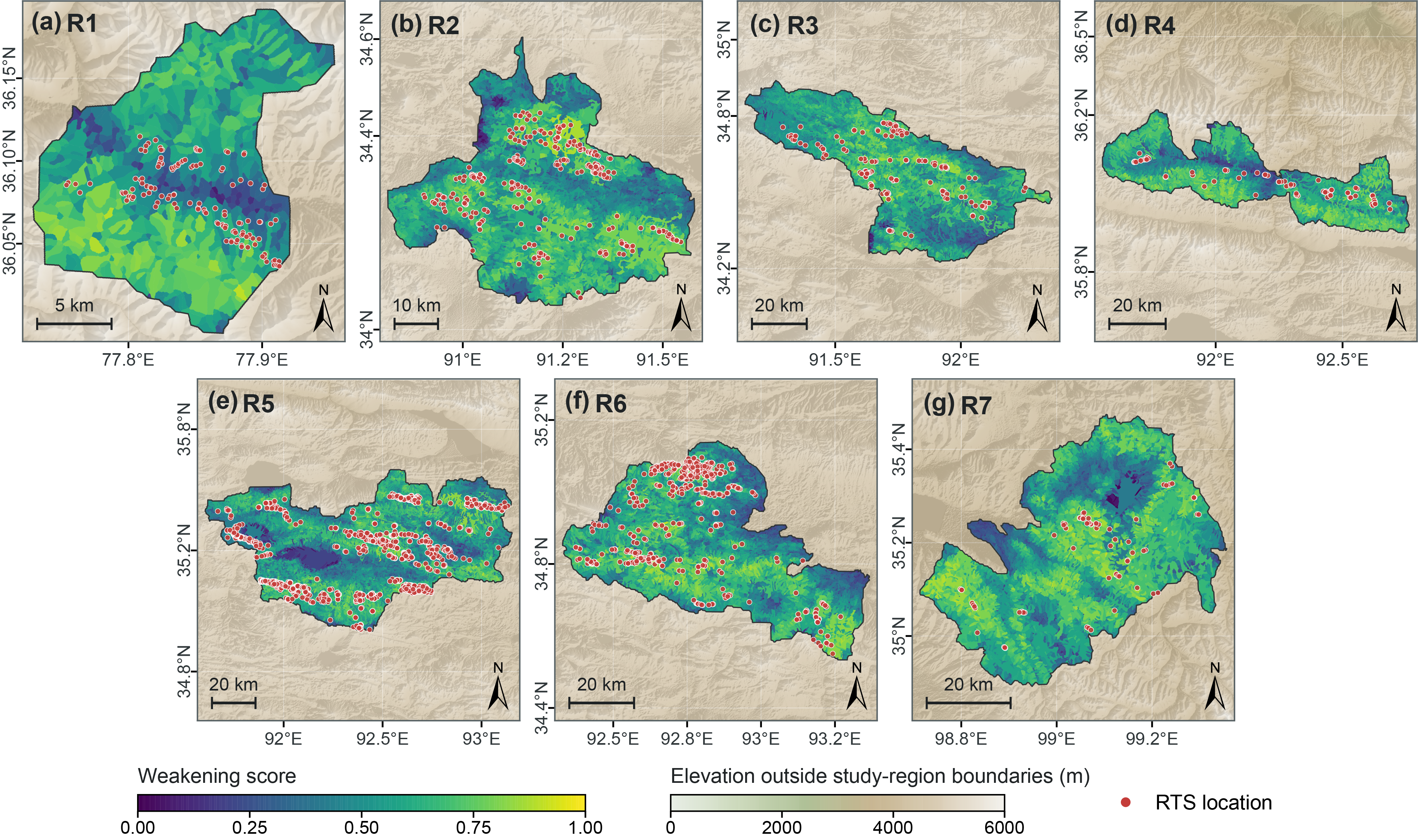}

\caption{Hillslope Weakening Fields constructed under complete leave-one-region-out evaluation. a--g, HWFs constructed for R1--R7 using the leave-one-region-out design, overlaid with RTS representative points matched to slope units.}
\label{fig:6}
\end{figure}

Across the seven independent evaluations, RTSs preferentially occurred on hillslopes with higher weakening scores in most regions (Fig.~\ref{fig:6}). We quantified this correspondence by comparing weakening scores between RTS-bearing and RTS-unrecorded slope units.

\begin{figure}[htbp]
\centering
\includegraphics[width=0.98\textwidth,height=0.56\textheight,keepaspectratio]{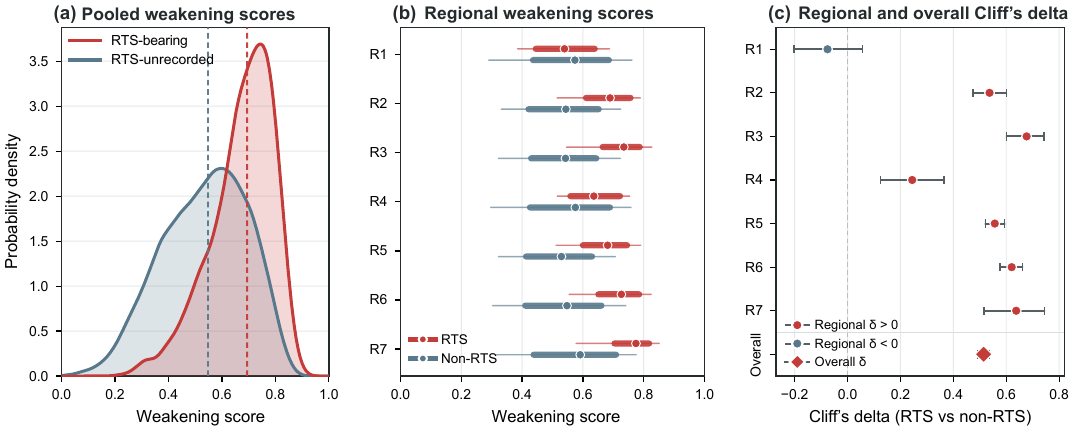}

\caption{RTS-bearing slope units have higher weakening scores in most independent evaluation regions. a, Probability density distributions of weakening scores for RTS-bearing and RTS-unrecorded slope units pooled across the seven study regions. The two density distributions are normalized separately, and dashed lines indicate the median of each group. b, Distributions of weakening scores for RTS-bearing and RTS-unrecorded slope units in R1--R7. Thin lines indicate the 10th--90th percentile range, thick lines indicate the interquartile range, and points indicate the median. c, Cliff's delta for R1--R7 and for all regions pooled. Horizontal lines indicate 95\% confidence intervals from 999 bootstrap resamples. Circles indicate Cliff's delta for individual regions, and the diamond indicates the pooled Cliff's delta across all regions. Positive values indicate that RTS-bearing slope units generally have higher weakening scores than RTS-unrecorded slope units.}
\label{fig:7}
\end{figure}

Considering all seven study regions collectively, the score distributions of RTS-bearing and RTS-unrecorded slope units overlapped, but the distribution for RTS-bearing units was clearly shifted toward higher values (Fig.~\ref{fig:7}a). The median weakening score was 0.694 for RTS-bearing slope units and 0.548 for RTS-unrecorded slope units, giving a median difference of 0.146. The overall Cliff's delta was 0.514, with a bootstrap 95\% confidence interval of 0.491--0.538, indicating a strong and stable tendency for a randomly selected RTS-bearing slope unit to have a higher weakening score than a randomly selected RTS-unrecorded slope unit.

Regional results further showed that this overall correspondence was not confined to only a few study regions (Fig.~\ref{fig:7}b, c). In R2, R3, R5, R6, and R7, the weakening-score distributions of RTS-bearing slope units were clearly shifted toward higher values relative to those of RTS-unrecorded slope units. Across these five regions, Cliff's delta ranged from 0.537 to 0.677, and all bootstrap 95\% confidence intervals lay entirely above zero. A positive but weaker correspondence was also observed in R4 (Cliff's delta = 0.244, 95\% confidence interval = 0.125--0.364). In R1, the difference was slightly negative (Cliff's delta = -0.076), with the confidence interval crossing zero (-0.203--0.056), indicating no clear positive HWF--RTS correspondence. Overall, positive correspondence was observed in six of the seven regions, with consistently strong relationships in R2, R3, R5, R6, and R7.

Across most independent evaluation regions, RTSs preferentially occurred on hillslopes with higher weakening scores within the HWF. This hillslope-level correspondence characterizes the localization of RTSs on higher-weakening hillslopes, whereas their spatial clustering requires consideration of the broader spatial structure of weakening. Accordingly, we next examine whether RTSs correspond to spatially continuous high-weakening domains in the HWF and to elevated weakening in their surrounding neighborhoods.

\subsection{RTS localization and clustering correspond to spatially extensive and locally continuous high-weakening domains in the HWF}\label{sec:2-4}

\begin{figure}[htbp]
\centering
\includegraphics[width=0.98\textwidth,height=0.46\textheight,keepaspectratio]{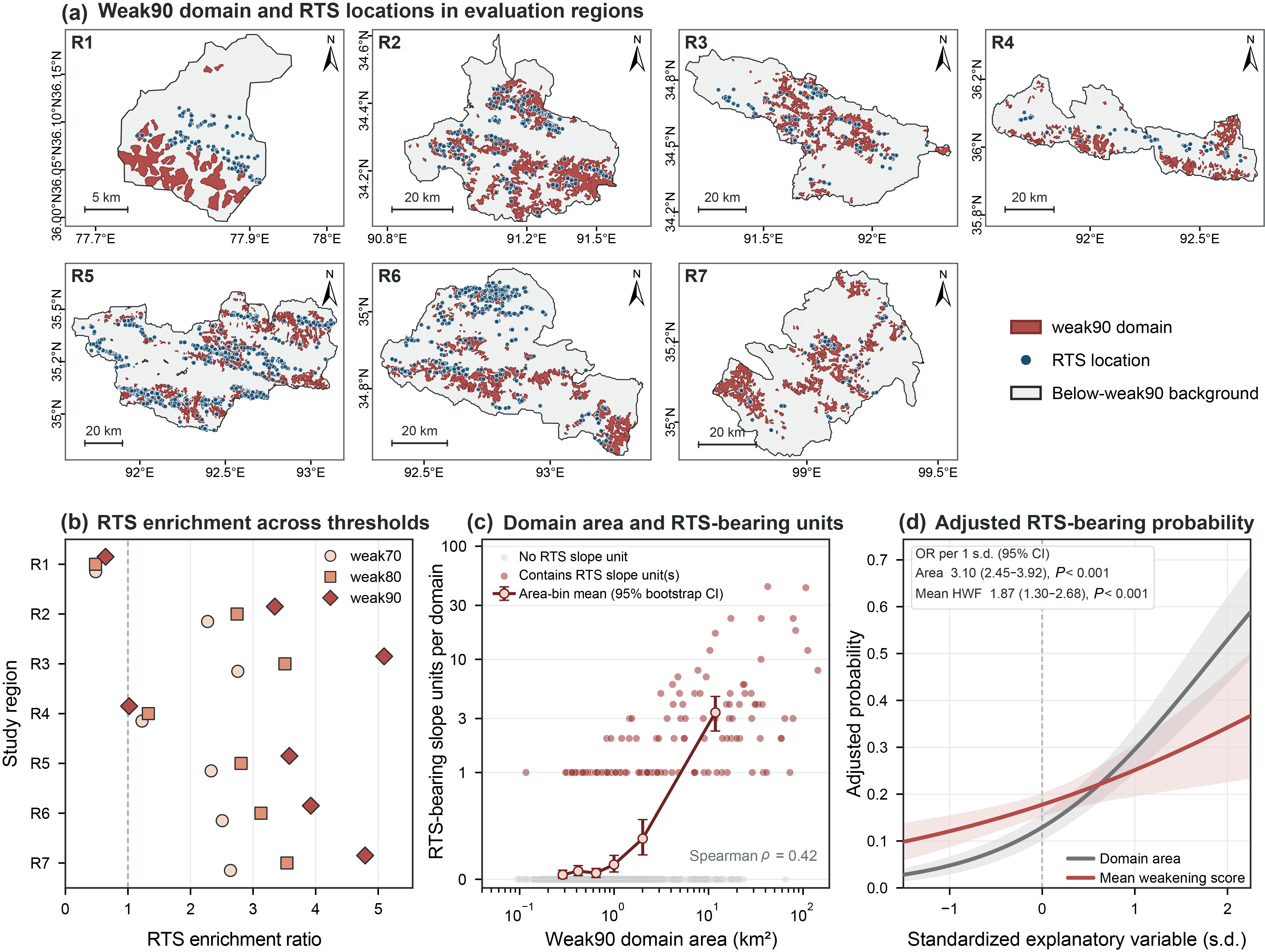}
\captionsetup{font=footnotesize}
\caption{High-weakening domains in the HWF correspond to the selective distribution of RTSs. a, Spatial distribution of weak90 high-weakening domains in each study region. Weak90 denotes the approximately top 10\% of slope units ranked by weakening score within each region. Adjacent weak90 slope units connected by shared boundaries were grouped into connected high-weakening domains. RTS representative points are overlaid. b, RTS enrichment within the weak70, weak80, and weak90 high-weakening areas, using the proportion of RTS-bearing slope units among all valid slope units in the corresponding region as the regional background. c, Relationship between weak90 high-weakening domain area and the number of RTS-bearing slope units within each domain. High-weakening domains are shown in gray or dark red according to whether they contain RTS-bearing slope units. Dark red points and lines show the mean number of RTS-bearing slope units and the corresponding 95\% bootstrap confidence intervals for six area-based groups of all weak90 high-weakening domains. Confidence intervals were obtained from 999 bootstrap resamples with replacement within each group. d, Relationships of weak90 high-weakening domain area and mean weakening level with RTS presence. The two curves show how the estimated probability that a domain contains at least one RTS-bearing slope unit changes as domain area increases or as the mean weakening score increases. Probabilities were estimated using a binomial generalized linear model including standardized log10-transformed domain area, standardized mean weakening score, and study region. For each curve, only the focal variable was varied, while the observed values of the other variable and the original regional composition were retained, and predicted probabilities were averaged across all weak90 high-weakening domains. Shaded areas indicate 95\% confidence intervals derived from the HC1 robust covariance matrix and the delta method. Insets report the odds ratio, HC1 robust 95\% confidence interval, and two-sided P value for each 1-SD increase in the corresponding variable.}
\label{fig:8}
\end{figure}

We next examined whether RTS localization and clustering correspond not only to higher weakening levels on individual hillslopes but also to spatially extensive and continuous high-weakening states. To represent different levels of high weakening, slope units within the top 30\%, 20\%, and 10\% of weakening scores in each region were defined as weak70, weak80, and weak90, respectively. High-weakening domains were then defined as connected areas formed by adjacent slope units above a given threshold. The spatial overlap of weak90 domains with RTS representative points indicates that, in most regions, RTSs are concentrated within a limited number of high-weakening domains (Fig.~\ref{fig:8}a).

RTS enrichment across different weakening thresholds further shows that RTS spatial localization becomes stronger as hillslope weakening increases. In R2, R3, R5, R6, and R7, the RTS occurrence rate within weak70 areas was 2.27--2.76 times the corresponding regional background rate. This enrichment increased to 2.74--3.54 in weak80 and further to 3.35--5.09 in weak90 (Fig.~\ref{fig:8}b). As the weakening threshold increases, the spatial extent of high-weakening areas becomes smaller, but the enrichment of RTSs relative to the regional background generally becomes stronger. This pattern indicates that RTSs become increasingly concentrated within a limited set of more strongly weakened hillslopes and high-weakening areas.

The correspondence between RTSs and the HWF is also reflected in the spatial extent of high-weakening domains. High-weakening domain area was positively correlated with the number of RTS-bearing slope units within each domain (Spearman's $\rho$ = 0.42, P < 0.001). Most small domains lacked recorded RTSs, whereas larger domains generally contained more RTSs (Fig.~\ref{fig:8}c). When domain extent and mean weakening level were considered together while accounting for regional differences, binary generalized linear modeling showed that both domain properties were positively associated with RTS presence: for each 1-SD increase in log10-transformed domain area, the odds ratio of containing RTSs was 3.10 (95\% CI, 2.45--3.92; P < 0.001); for each 1-SD increase in mean domain weakening score, the odds ratio was 1.87 (95\% CI, 1.30--2.68; P < 0.001) (Fig.~\ref{fig:8}d). Thus, high-weakening domains with recorded RTSs were characterized by both higher overall weakening levels and greater spatial extent.

\begin{figure}[htbp]
\centering
\includegraphics[width=0.98\textwidth,height=0.56\textheight,keepaspectratio]{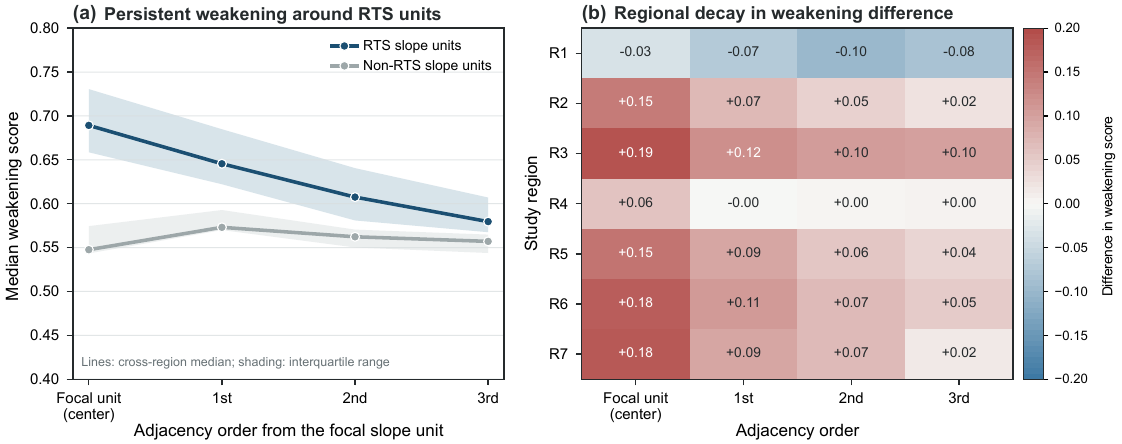}

\caption{Weakening-score differences between RTS-bearing and RTS-unrecorded slope units extend into surrounding neighborhoods. a, Median weakening scores of RTS-bearing slope units and all RTS-unrecorded slope units, together with the median scores of their first-, second-, and third-order adjacency rings. Center denotes the slope units themselves: the blue curve represents RTS-bearing slope units, and the gray curve represents RTS-unrecorded slope units. Solid lines show the median of the seven regional medians, and shaded areas indicate the interquartile range across regions. b, Differences in weakening scores between the RTS-bearing group and the RTS-unrecorded group in each study region. The difference at center is the difference between the median weakening scores of RTS-bearing slope units and all RTS-unrecorded slope units. The differences at the first-, second-, and third-order positions are the differences between the median weakening scores of the corresponding adjacency rings of the two groups. Positive values indicate that RTS-bearing slope units or their adjacency rings have higher weakening scores than the corresponding RTS-unrecorded reference group.}
\label{fig:9}
\end{figure}

Neighborhood analysis further showed that the higher weakening states associated with RTSs were not confined to RTS-bearing hillslopes, but extended into surrounding hillslopes. The across-region median weakening score was 0.689 for RTS-bearing slope units and 0.548 for RTS-unrecorded slope units. Moving outward from the center slope unit, the median neighborhood score declined from 0.646 in the first-order neighborhood to 0.608 in the second-order neighborhood and 0.580 in the third-order neighborhood. The corresponding values around RTS-unrecorded slope units were 0.573, 0.562, and 0.557, respectively (Fig.~\ref{fig:9}a). The difference between the two across-region median curves decreased from 0.141 at the center slope unit to 0.073, 0.046, and 0.023 at the first-, second-, and third-order neighborhoods, respectively, but remained positive through the third order, indicating that RTSs more often correspond to spatially continuous higher-weakening environments than to isolated slope units with high weakening scores.

The same neighborhood pattern was evident across most study regions. In R2, R3, R5, R6, and R7, weakening scores remained higher for RTS-bearing slope units and their surrounding hillslopes than for RTS-unrecorded slope units and their surroundings, extending through the third-order neighborhood. Even at the third order, the weakening-score differences remained positive, ranging from 0.016 to 0.100 (Fig.~\ref{fig:9}b). Although the magnitude of this contrast varied among regions, the difference was consistently greatest at the RTS-bearing slope unit itself and generally diminished toward the surrounding hillslopes.

RTS enrichment within high-weakening areas of the HWF, the spatial extent of high-weakening domains, and the persistent neighborhood differences in weakening scores show that RTS spatial localization and clustering correspond not only to higher weakening levels on individual hillslopes, but also to high-weakening domains characterized by greater spatial extent and continuity. Therefore, RTS localization and clustering correspond to both the degree and spatial properties of hillslope weakening states.

\subsection{HWF--RTS correspondence shows clear cross-regional repeatability}\label{sec:2-5}

The complete leave-one-region-out evaluation showed positive HWF--RTS correspondence in six of the seven regions, with consistently strong relationships in R2, R3, R5, R6, and R7. In these five regions, RTSs preferentially occurred on hillslopes with higher weakening scores (Section 2.3) and remained enriched within the highest-weakening areas (Section 2.4). Within the top 10\% of weakening scores, RTS occurrence rates were 3.35--5.09 times the corresponding regional background rates (Fig.~\ref{fig:8}b). This repeated pattern indicates that the association between RTS spatial selectivity and hillslope weakening state recurs across independent permafrost landscapes.

This cross-regional repeatability is evident not only in the preferential occurrence of RTSs on higher-weakening hillslopes, but also in the spatial continuity and connectivity of high-weakening states and their correspondence with RTS clustering. High-weakening hillslopes formed spatially continuous and connected high-weakening domains (Fig.~\ref{fig:5}). RTSs were not randomly distributed among these hillslopes, but were more concentrated within high-weakening domains and their surrounding continuous weakening neighborhoods. Larger high-weakening domains and domains with higher mean weakening scores were more likely to contain RTSs, while higher weakening states around RTS-bearing hillslopes gradually declined from the center toward surrounding neighborhoods (Fig.~\ref{fig:8} and 9). Overall, HWF--RTS correspondence was consistently observed across most study regions under the leave-one-region-out evaluation, from higher weakening levels on individual hillslopes to the spatial continuity of high-weakening areas and the selective distribution of RTSs within these areas.

Overall, HWF--RTS correspondence recurred across most regions, although its strength varied. Positive HWF--RTS correspondence was not supported in R1; the relationship was weaker in R4 but remained consistently strong in R2, R3, R5, R6, and R7. These regional differences suggest that HWF--RTS correspondence is related not only to regional environmental background but also to whether the environmental factors associated with RTS spatial selectivity retain broadly similar relative importance across regions.

\section{Discussion}\label{sec:3}

\subsection{How broad permafrost change is translated into localized RTS responses}\label{sec:3-1}

Our results address the spatial mismatch between widespread changes in permafrost environments and highly localized RTS responses. Permafrost warming and degradation provide a broad regional background for RTS development, but these changes are not expressed uniformly as hillslope instability \citep{ref20,ref21,ref22}. Spatial differences in thermal state, moisture conditions, topographic position, permafrost and ground ice, and land cover jointly shape contrasting hillslope weakening states \citep{ref23,ref24}. RTSs preferentially occur on more strongly weakened hillslopes. Differences in weakening state among hillslopes therefore provide a spatial link between broad environmental change and localized geomorphic responses.

The HWF does not represent any single environmental condition, but rather the hillslope weakening state jointly shaped by multiple conditions. Ground ice provides the material basis for RTS development; thermal state affects permafrost stability and ice thaw; moisture influences heat transfer, soil softening, and material transport; and topography constrains gravitational release, water movement, and toe-boundary conditions \citep{ref03,ref05,ref12,ref25}. These conditions interact with and constrain one another, and their effects may also involve thresholds, preferred ranges, and nonmonotonic responses \citep{ref24}. Similar values of a single environmental variable therefore do not necessarily correspond to similar hillslope weakening states \citep{ref26}. RTS spatial selectivity is therefore better understood in relation to hillslope weakening states jointly shaped by multiple conditions than to any single environmental variable considered separately.

The recurrence of HWF--RTS correspondence under complete leave-one-region-out evaluation further indicates that this relationship is not confined to a single regional setting. RTS-bearing slope units generally retained higher weakening scores across most independent evaluation regions, supporting the broader relevance of hillslope weakening states to RTS spatial selectivity. These results suggest that widespread changes in permafrost environments are translated into localized geomorphic responses through spatial differentiation of hillslope weakening states.

\subsection{Spatial properties of the HWF explain the localized and clustered distribution of RTSs}\label{sec:3-2}

Understanding why RTSs are not only localized but also spatially clustered requires consideration of the spatial properties of hillslope weakening states. Our results show that high-weakening hillslopes are not distributed in isolation. Instead, they form connected high-weakening domains that extend across multiple neighboring hillslopes and over characteristic spatial distances. RTSs preferentially occur within these high-weakening domains and their surrounding continuous weakening neighborhoods. RTS localization and clustering are therefore better understood in relation to the spatial properties of hillslope weakening states than to individual environmental conditions considered separately.

The HWF is characterized by the degree of weakening, spatial extent, and neighborhood continuity. The weakening score of a slope unit represents its relative weakening state jointly formed by multiple environmental conditions. The spatial correlation length describes how far similar weakening states extend, while the connectivity of high-weakening slope units characterizes whether these hillslopes remain spatially dispersed or form continuous high-weakening domains. Neighborhood analysis further shows that the higher weakening states associated with RTS-bearing slope units are not restricted to RTS-bearing hillslopes, but extend into surrounding hillslopes and gradually decline with increasing neighborhood distance. This pattern is consistent with the tendency for topographic position, hydrological conditions, surface energy conditions, and permafrost conditions to vary continuously across individual slope-unit boundaries \citep{ref27}. RTSs therefore correspond not simply to isolated high-weakening hillslopes, but to spatially continuous weakening environments extending across multiple hillslopes.

The spatial extent and mean weakening level of high-weakening domains describe two important properties of these domains. Larger domains indicate that environmental conditions favorable for hillslope weakening persist continuously across a greater area, whereas higher mean weakening scores indicate stronger overall weakening states within the domain. RTSs are therefore associated not only with weakening levels on individual hillslopes, but also with the spatial extent and connectivity of high-weakening states. This result is consistent with studies of spatial heterogeneity in hillslope strength, which show that the magnitude, spatial scale, and connectivity of local weak zones can jointly influence the spatial pattern of slope failure \citep{ref18,ref28}. The HWF extends this spatial-field perspective to RTS localization and clustering by representing hillslope weakening as a spatially structured state jointly shaped by multiple environmental conditions across permafrost landscapes.

Although HWF--RTS correspondence recurred across most regions, its strength varied among regions. R2, R3, R5, R6, and R7 retained consistently strong positive relationships, whereas the correspondence was weaker in R4 and was not supported in R1. Further regional comparisons suggest that these differences are not readily explained by environmental background alone. R1 differs broadly from the other regions in its environmental characteristics, while R4 and R7 also differ in several thermal, moisture, topographic, and permafrost conditions. However, the environmental factors associated with RTS spatial selectivity and their relative importance in R7 remain broadly similar to those in the other regions, and strong positive HWF--RTS correspondence is retained. This contrast suggests that regional environmental dissimilarity alone does not account for variation in HWF--RTS correspondence. Cross-regional similarity in the relative importance of environmental factors associated with RTS spatial selectivity may also contribute to the persistence of this correspondence. Detailed analyses of regional environmental distributions and environmental-factor importance rankings are provided in Supplementary Section S3 and Supplementary Figs. S3--S7.

The presence of hillslopes with high weakening scores but no recorded RTSs indicates that long-term weakening does not imply that an RTS must already have occurred. Short-term processes, including local exposure of ground ice, toe erosion, and anomalous hydrothermal events, may influence whether and where RTSs develop \citep{ref29}. The temporal coverage and completeness of RTS inventories may further affect whether such events are recorded \citep{ref30}. The HWF should therefore be interpreted as a representation of long-term hillslope weakening states and their spatial structure, rather than as a deterministic threshold for RTS occurrence. Distinguishing long-term hillslope preconditioning from short-term triggering will require RTS observations with explicit temporal information.

\subsection{Implications for spatially heterogeneous responses of permafrost landscapes}\label{sec:3-3}

Our results provide a basis for understanding why permafrost landscapes can exhibit spatially heterogeneous geomorphic responses under widespread warming and degradation. Permafrost warming and degradation may affect large areas, but the intense hillslope disturbance represented by RTSs is concentrated on only a limited proportion of permafrost hillslopes \citep{ref31,ref32}. This spatial selectivity is better understood in relation to spatial heterogeneity in multiple environmental conditions, including thermal state, moisture conditions, topography, permafrost, and ground ice, than to any single environmental condition considered separately. These environmental differences give rise to contrasting hillslope weakening states that can form connected high-weakening domains across the landscape. RTSs are preferentially concentrated within these domains. Widespread changes in permafrost environments therefore do not produce spatially uniform geomorphic responses, but are translated into highly selective hillslope instability through the spatial differentiation and connectivity of hillslope weakening states.

The spatial selectivity of RTSs also suggests that the resulting geomorphic disturbance and material release may be spatially concentrated. RTSs can mobilize large volumes of permafrost material, sediment, and organic carbon and can affect downstream hydrological and ecological processes \citep{ref06,ref33,ref34}. Areas with concentrated RTS activity may therefore become important source areas of sediment, organic carbon, and associated hydrological disturbance. Although sediment transport, carbon release, and downstream ecological effects were not quantified here, the observed concentration of RTSs suggests that these impacts may also be concentrated within a limited number of high-weakening hillslopes and connected domains. This spatial concentration suggests a potential basis for similarly heterogeneous patterns of material transport and ecohydrological impacts associated with permafrost degradation.

These implications may extend beyond RTSs. Active-layer detachments, permafrost riverbank failures, thaw erosion, and other thermokarst processes differ in material conditions, topographic settings, and process mechanisms, but they also occur selectively within broadly degrading permafrost landscapes \citep{ref31,ref35,ref36}. This similarity raises a broader possibility that spatially heterogeneous geomorphic responses may arise where regional environmental change interacts with spatially structured local conditions. From this perspective, broad climatic and permafrost changes may be filtered by local environmental heterogeneity before being expressed as discrete geomorphic disturbances. The HWF therefore provides a conceptual basis for considering how spatially structured environmental preconditioning may contribute to heterogeneous geomorphic responses across permafrost landscapes.

For practical applications, this spatial selectivity also suggests that regional investigation and monitoring could be prioritized toward hillslopes with higher weakening scores and connected high-weakening domains. Such areas could be prioritized for field surveys, UAV mapping, geophysical investigations, and deformation monitoring, particularly for broad-scale screening along transportation and other linear infrastructure corridors on the Tibetan Plateau \citep{ref37,ref38}. The HWF cannot replace site-scale investigations of ground ice, deformation monitoring, or slope-stability assessment. It can, however, help narrow the areas requiring detailed investigation and direct limited monitoring resources toward hillslopes and connected domains where multiple weakening conditions coincide.

\subsection{Limitations and future work}\label{sec:3-4}

Although the complete leave-one-region-out design separates the identification of environmental weakening-response relationships from the independent evaluation of HWF--RTS correspondence, all seven study regions are located on the Tibetan Plateau. Whether these relationships remain applicable to permafrost landscapes beyond the plateau therefore requires further testing. Second, the current RTS inventory lacks consistent information on occurrence timing, limiting our ability to separate long-term hillslope preconditioning from short-term triggering processes. The current HWF primarily represents long-term hillslope weakening states and does not explicitly represent short-term triggers such as extreme rainfall, concentrated snowmelt, erosion, or anomalous warming. Available data on ground ice, active-layer conditions, and freeze--thaw processes are also limited by spatial resolution and proxy uncertainty, while slope-unit averages may mask localized high-weakening conditions within individual units. In addition, riverbank and lakeshore erosion, road disturbance, lithology, sediment properties, and extreme events are not yet fully represented. These missing factors may help explain why some RTSs occur on hillslopes with relatively low weakening scores, while other hillslopes with high scores contain no recorded RTSs.

Future work should test the HWF using independent regions beyond the Tibetan Plateau and multitemporal RTS inventories with explicit occurrence timing. Such data would allow testing whether the HWF framework can be extended to characterize temporal changes in hillslope weakening states and their relationship with RTS development. High-resolution remote sensing, borehole observations, geophysical surveys, InSAR, and more detailed ground-ice data could support multiscale representations of weakening states that retain sub-slope spatial heterogeneity while preserving the hillslope-scale framework, allowing direct tests of whether localized high-weakening conditions are obscured by slope-unit averaging. Future analyses should also incorporate erosional boundaries, human disturbance, and extreme hydrothermal events to distinguish long-term weakening conditions from short-term triggering processes. These developments would provide a stronger basis for evaluating the spatial and temporal applicability of the HWF under changing permafrost conditions.

Overall, our results show that RTS localization and clustering are associated with both the degree and spatial properties of hillslope weakening states. This correspondence recurs across most study regions under independent evaluation, although its strength varies among regions. Our findings provide a spatial explanation for how widespread permafrost change can be translated into highly localized and heterogeneous geomorphic responses across permafrost landscapes.

\section{Materials and methods}\label{sec:4}

\subsection{Study regions and RTS inventory}\label{sec:4-1}

We used seven study regions to examine the relationship between the HWF and RTS spatial selectivity while reducing the possibility that the identified environmental responses reflected only local statistical patterns within a single region. The seven study regions, denoted R1--R7, span different environmental settings across the Tibetan Plateau (Fig.~\ref{fig:3} and Supplementary Data S1). Study-region boundaries were defined from level-12 subbasins in HydroBASINS v1 \citep{ref19} and were used consistently in subsequent HWF construction and RTS spatial analyses.

RTS records were collected from several published or publicly available datasets, including the 2022 inventory derived from the annual 2016--2022 Tibetan Plateau RTS dataset \citep{ref08,ref39}, the 2021 RTS records from the 2008--2021 Beiluhe Basin inventory \citep{ref14,ref40}, RTS records from the Hoh Xil region for 2018--2019 \citep{ref41}, and the 2019 RTS inventory for the Qinghai--Tibet Engineering Corridor \citep{ref09,ref42}. These datasets were combined to construct an integrated RTS inventory. After all inventories were transformed to a common coordinate reference system and clipped to the study regions, spatial duplicates were removed sequentially, with the 2022 plateau-wide inventory used as the reference. Supplementary RTS polygons that intersected, overlapped with, or touched the boundary of an existing RTS polygon were treated as potential duplicates, and only polygons spatially disjoint from the existing inventory were retained. The final integrated inventory contained 2,672 RTS polygons.

To assign each RTS a spatial location representative of its hillslope setting, we generated DEM-constrained upper-slope representative points rather than directly using RTS polygon centroids. For each RTS polygon, the upper portion was identified using the 90th percentile of DEM elevations within the polygon, and the centroid of this upper area was used as the representative point. If the centroid fell outside the corresponding area, an interior representative point was used instead. If no valid DEM pixels were available within an RTS polygon, an interior point of the original polygon was used as a fallback. One representative point was generated for each RTS polygon.

HWF construction and spatial analyses were conducted using geomorphic slope units as the basic analytical units. RTSs are hillslope-scale geomorphic processes involving headwall retreat, thaw, and material transport. Slope units allow thermal, hydrological, topographic, permafrost, and ground-ice conditions to be summarized over geomorphically meaningful hillslopes, while their shared boundaries provide a natural topological framework for analyzing spatial connectivity, domain structure, and neighborhood relationships of weakening states within the HWF. Slope units were independently delineated for each study region from the ~30-m SRTM GL1 DEM using ``r.slopeunits'' in GRASS GIS 8.4 \citep{ref43}. A total of 37,749 slope units were obtained across the seven regions, with regional counts provided in Supplementary Data S1.

RTS upper-slope representative points were assigned to slope units by spatial intersection. Of the 2,672 RTS representative points, 2,667 were successfully matched to slope units, while five fell outside the slope-unit analysis extent. A slope unit containing at least one RTS representative point was defined as an RTS-bearing slope unit, and all remaining units were defined as RTS-unrecorded slope units. This resulted in 1,342 RTS-bearing and 36,407 RTS-unrecorded slope units. Slope-unit-based analyses used binary RTS occurrence status, defined by whether a slope unit contained at least one RTS representative point, thereby avoiding repeated weighting of multiple RTS objects occurring within the same slope unit.

\subsection{Acquisition and processing of environmental data}\label{sec:4-2}

We integrated multiple sources of environmental information that characterize long-term environmental preconditioning of hillslopes, including information on thermal state, hydrology, terrain, permafrost, and surface cover. Together, these variables characterize hillslope thermal and freeze--thaw conditions, water availability and storage, topographic and hydrogeomorphic setting, permafrost and ground-ice conditions, and land-surface cover relevant to RTS development.

Because the environmental datasets differed in spatial resolution and data format, all datasets were transformed to a common spatial reference and summarized for each slope unit. Continuous raster variables were generally represented by the mean of valid pixels within each slope unit, while distance-related variables were summarized according to their spatial definitions. Categorical variables were represented using the dominant class and its corresponding code. This procedure produced 50 source environmental variables. Their definitions, temporal coverage, original spatial resolutions, data sources, slope-unit aggregation methods, data completeness, and quality-control procedures are provided in Supplementary Data S2.

\subsection{Construction and evaluation of the Hillslope Weakening Field (HWF)}\label{sec:4-3}

\subsubsection{Construction of the Hillslope Weakening Field}\label{sec:4-3-1}

To characterize the hillslope weakening states and the spatial properties of these states, we developed the Hillslope Weakening Field (HWF). The HWF represents relative differences in hillslope weakening states across the landscape and provides a framework for testing whether the spatial properties of these states help explain RTS spatial selectivity.

The construction of HWF is composed of four steps. (1) Numerical environmental factors (i.e., all factors except the categorical land-use factor) were transformed into within-region percentile ranks to place variables with different units on a common relative scale. (2) RTS records from the reference regions is used to transform each environmental factor into a relative weakening score ranging from 0 to 1 according to its relationship with RTS occurrence. (3) Relative evidence weights are assigned to environmental factors according to response strength, response-pattern support, area robustness, data completeness, and variable redundancy. (4) Factor-level weakening scores are combined by weighted aggregation for each slope unit, and the spatial distribution of the resulting slope-unit weakening scores constitutes the HWF. Detailed procedures for within-region percentile transformation, factor-level weakening-score derivation, evidence-weight determination, and weighted aggregation are provided in Supplementary Section S1.

The identification of environmental weakening responses and their relative evidence weights relies on relationships between recorded RTSs and environmental conditions. HWF--RTS correspondence therefore requires evaluation using RTS records independent of that used for HWF construction. We accordingly adopted a complete leave-one-region-out design to separate HWF construction from HWF--RTS evaluation at the regional level, as described in Section 4.3.2.

\subsubsection{Independent evaluation of HWF--RTS correspondence}\label{sec:4-3-2}

We adopted seven leave-one-region-out evaluations to examine HWF--RTS correspondence. In each evaluation, one of R1--R7 was designated as the independent evaluation region, while the remaining six regions served as reference regions. Using the six reference regions, the complete HWF construction procedure described in Section 4.3.1 was repeated to derive a fixed set of HWF construction rules for that evaluation. The established rules were then applied to the environmental data of the independent evaluation region to derive slope-unit weakening scores and construct the regional HWF. RTS records from the evaluation region was introduced only after HWF construction to evaluate HWF--RTS correspondence. Repeating this procedure for all seven regions produced HWFs independently evaluated across all the seven study regions.

By excluding the RTS records of the evaluation region when determining the HWF construction rules, the same RTS records were not used in both HWF construction and evaluation. This design tests whether environmental weakening relationships identified from the six reference regions remain associated with RTS spatial selectivity in an independent region. Leaving out an entire region also provides a more rigorous evaluation for data with spatial dependence and regional structure than randomly dividing individual spatial units \citep{ref44}. The complete leave-one-region-out design was therefore used to evaluate the cross-regional repeatability of HWF--RTS correspondence rather than predictive performance.

We conducted three analytical procedures to evaluate HWF--RTS correspondence using the complete leave-one-region-out design. First, we analyzed RTS spatial selectivity by quantifying RTS localization and nonrandom clustering, thereby defining the basic spatial phenomenon that the HWF is expected to explain. Second, without using RTS locations or occurrence status, we analyzed the spatial-field properties of the HWF to determine whether it exhibits spatial autocorrelation and spatially extensive and connected high-weakening domains. Third, within each independent evaluation region, we analyzed the HWF--RTS correspondence, including differences in weakening scores between RTS-bearing and RTS-unrecorded slope units, RTS enrichment within high-weakening areas, and correspondence involving high-weakening domains and their surrounding neighborhoods. These three analytical procedures are described in detail in Supplementary Section S2.

\section*{Acknowledgments}

\textbf{Funding:} This research was jointly supported by the National Science Foundation of China (Grants No. 42277161).

\end{document}